\pdfoutput=1

\documentclass[3p,times]{elsarticle}

\usepackage{ecrc}

\volume{00}

\firstpage{1}

\journalname{Nuclear Physics A}

\runauth{M. Verweij}

\jid{nupha}

\usepackage{amssymb}

\usepackage{subfigure}

\usepackage[figuresright]{rotating}

\begin{document}

\begin{frontmatter}



\dochead{}

\title{Comparison of radiative energy loss models in a hot QCD medium}


\author{Marta Verweij\fnref{fn1}}
\address[fn1]{Utrecht University}
\ead{m.verweij@uu.nl}

\begin{abstract}
  The suppression of high $p_{T}$ hadron production in heavy ion
  collisions is thought to be due to energy loss by gluon radiation off hard
  partons in a QCD medium. Existing models of QCD radiative energy loss in a
  color-charged medium give estimates of the coupling strength of the parton
  to the medium which differ by a factor of $5$. We will present a side-by-side
  comparison of two different formalisms to calculate the energy loss of light
  quarks and gluons: the multiple soft scattering approximation (ASW-MS) and
  the opacity expansion formalism (ASW-SH and WHDG-rad). A common
  time-temperature profile is used to characterize the medium. The results are
  compared to the single hadron suppression $R_{AA}$ at RHIC energies.

  In addition the influcence of homogeneous and non-homogeneous distribution
  of scattering centers is discussed. We find that using an equivalent brick
  overestimates the energy loss for long parton trajectories.

\end{abstract}


\end{frontmatter}




In these proceedings we present a comparison between radiative energy loss
models: the multiple soft scattering approximation ASW-MS
\cite{Salgado:2003gb}, and two opacity expansion formalisms ASW-SH
\cite{Salgado:2003gb} and (D)GLV \cite{Gyulassy:1999zd,Wicks:2005gt}.

\section{Treatment of medium geometry}
We use the medium temperature $T$ as a common variable for the different
formalisms and translate this to the input parameters needed to calculate the
energy loss. The participant density which is deduced from a Glauber
calculation is assumed to be proportional to $T^{3}$. We define the path
integral $J_{n}^0{(m)}$ from which we calculate the relevant parameters for
the two considered types of energy loss models:
\begin{equation}\label{eq:Jnm}
J_{n}^{(m)} = \int d\tau\;\tau^{n}\;T^{m}(\tau),
\end{equation}
where $\tau$ is the distance (time) along the parton trajectory. The density
of the medium starts expanding longitudinally with $1/\tau$ at formation time
$\tau_{0}=0.6$ fm. We assume a static density profile for times prior to
the formation time.
We assume that the parton does not interact with the medium after the
temperature is below freeze-out temperature which is set at $150$
MeV. Hadronization takes place in the vacuum.

\subsubsection{Multiple soft scattering approximation}
For the multiple soft scattering approximation $J_{0}^{3}$ and $J_{1}^{3}$ are
used in which $T^{3}$ is replaced by the local $\hat{q}(T)$ as is also done in
PQM \cite{Dainese:2004te}.  This provides an effective length and $\hat{q}$:
\begin{equation}\label{eq:effMS}
L_{eff} \propto \frac{2 J_{1}^{(3)}}{J_{0}^{(3)}} \:\: \mathrm{ and } \:\: \hat{q}_{eff} \propto \frac{J_{0}^{(3)} \cdot J_{0}^{(3)}}{2 J_{1}^{(3)}}.
\end{equation}

\subsubsection{Opacity expansion}
For the opacity expansion the input parameters to calculate the single
gluon radiation spectrum $dI/d\omega$ are:
\begin{equation}\label{eq:effGLV}
\bar{\omega_{c}} \propto T^{2}L \propto J_{0}^{(2)},\:\: \frac{L}{\lambda} \propto
TL \propto J_{0}^{(1)} \:\: \mathrm{ and} \:\: \bar{R} \propto \bar{\omega_{c}}^{2}L \propto T^{2}L^{2} \propto J_{0}^{(2)}\frac{J_{1}^{(m)}}{J_{0}^{(m)}}.
\end{equation}

The parton momentum distribution, calculated at leading order, is convoluted
with the energy loss probability distribution and the KKP fragmentation
function \cite{Kniehl:2000fe} to obtain the final charged hadron spectra. The
nuclear modification factor $R_{AA}$ is calculated for a wide range of values
of $\hat{q}$.  The best fit for each energy loss formalism is determined by
minimizing the modified $\chi^{2}$ \cite{Adare:2008cg,Adare:2008qa}. The
estimations of initial maximal $\hat{q}_{0}$ differ by a factor of $6$:
$20.3^{+0.6}_{-5.1}$ $GeV^{2}/\mathrm{fm}$ for ASW-MS, $5.7^{+0.3}_{-1.9}$
$GeV^{2}/\mathrm{fm}$ for (D)GLV and $3.2^{+0.3}_{-0.3}$ $GeV^{2}/\mathrm{fm}$
for ASW-SH.
\begin{figure}[!h]
\centering
\includegraphics[width=0.6\textwidth]{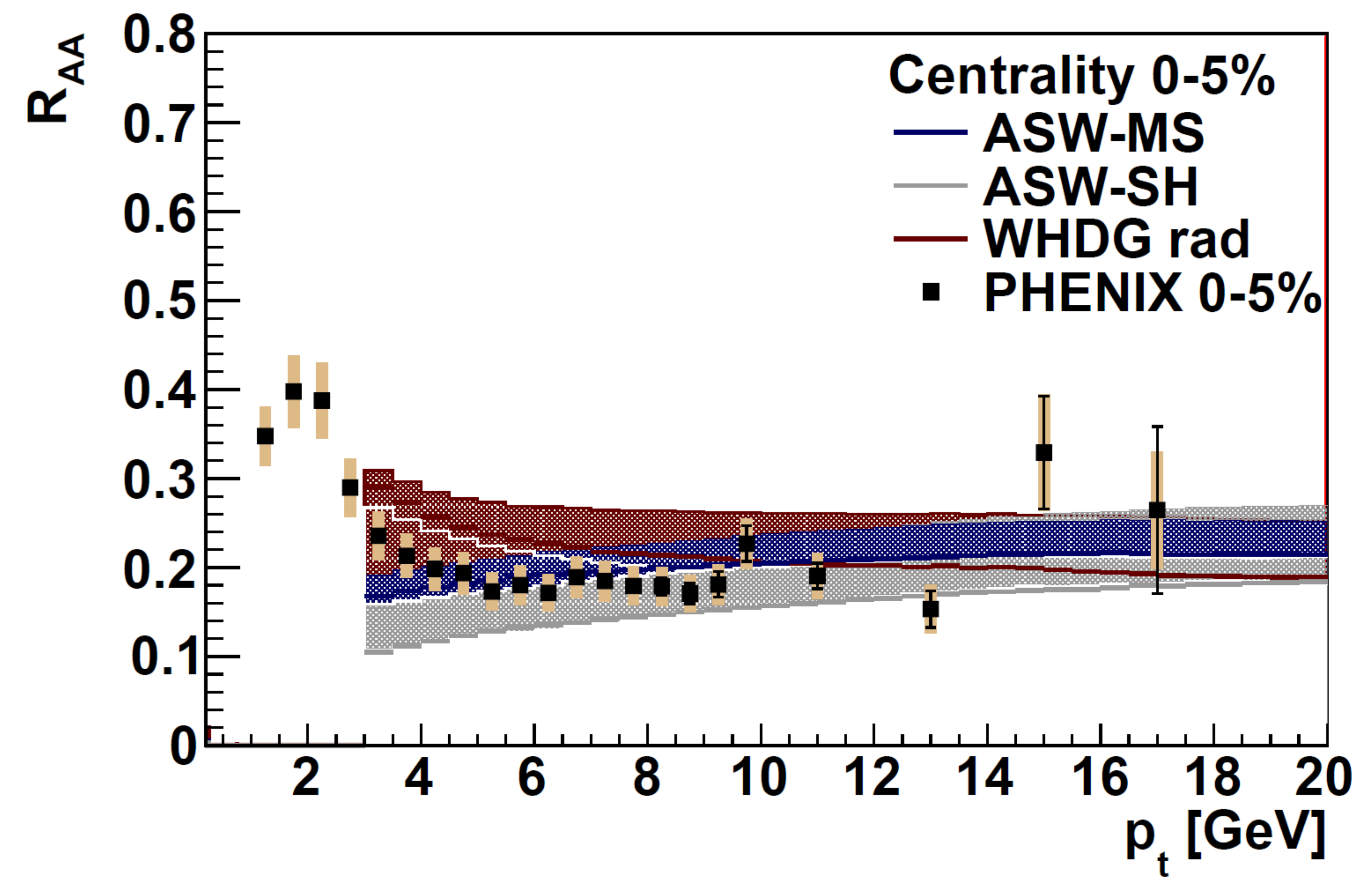}
\caption{Best model fits for measured nuclear modification factor $R_{AA}$
  \cite{Adare:2008cg} for the multiple soft scattering approximation and
  opacity expansion formalism. Shaded areas indicate the $1\sigma$ uncertainty
  on the fit \label{fig:RAAvsPtBestFits}}
\end{figure}

\section{Gluon emission at first order of opacity}

The medium-induced gluon emission spectrum at first order of opacity in the
GLV model \cite{Gyulassy:1999zd} can be written as \cite{CaronHuot:2010bp}:
\begin{equation}\label{eq:OESingleGluonSpec}
x\frac{dN_{g}}{dx} =
\frac{C_{R}C_{g}g^{2}}{2\pi^{3}}\int \frac{d^{2}\bf{q}}{(2\pi)^{2}}d^{2}{\bf
  k}\;dz\; C({\bf q},z) \times \mathcal{K}({\bf k},{\bf q}, z),
\end{equation}
in which ${\bf k}$ is the transerse momentum of the emitted gluon and ${\bf
  q}$ the transverse momentum exchanged with a scattering center in the
medium,
\begin{equation}\label{eq:Cqz}
C({\bf q},z) = \frac{1}{C_{R}}(2\pi)^{2}\frac{d^{2}\Gamma_{el}({\bf
    q},z)}{d^{2}{\bf q}}.
\end{equation}
and
\begin{equation}\label{eq:EnergyKernel}
\mathcal{K}({\bf k},{\bf q}, z) = \frac{{\bf{k \cdot q}} ({\bf{k}}-{\bf{q}})^{2}-\beta^{2}{\bf{q}}
  \cdot ({\bf k} -{\bf q})}{[({\bf k} - {\bf
      q})^{2}+\beta^{2}]^{2}({\bf k}^{2}+\beta^{2})} \times \left[1-\mathrm{cos}\left(\frac{({\bf k}-{\bf q})^{2}+\beta^{2}}{2Ex}z\right)\right].
\end{equation}
In the original GLV publications, the scattering rate $C({\bf q},z)$ is
separated into three terms: the number of scattering centers $L/\lambda$, a
normalised Yukawa potential and a normalised density profile $\rho(z)$.

The scattering rate per unit path length $\frac{d^{2}\Gamma_{el}({\bf
    q},z)}{d^{2}{\bf q}}$ carries all information about the density of the
medium and therefore depends on the local position $z$ of the parton in the
medium.

\subsection{Sensitivy to choice of density profile $\rho(z)$}
\begin{figure*}
  \centering
  \subfigure[Single gluon spectrum for GLV and WHDG
  implementation at same medium temperature. Normalized Yukawa potential and $\rho(z)$]{\label{fig:SingleGluonSpec}
    \includegraphics[width=0.35\textwidth]{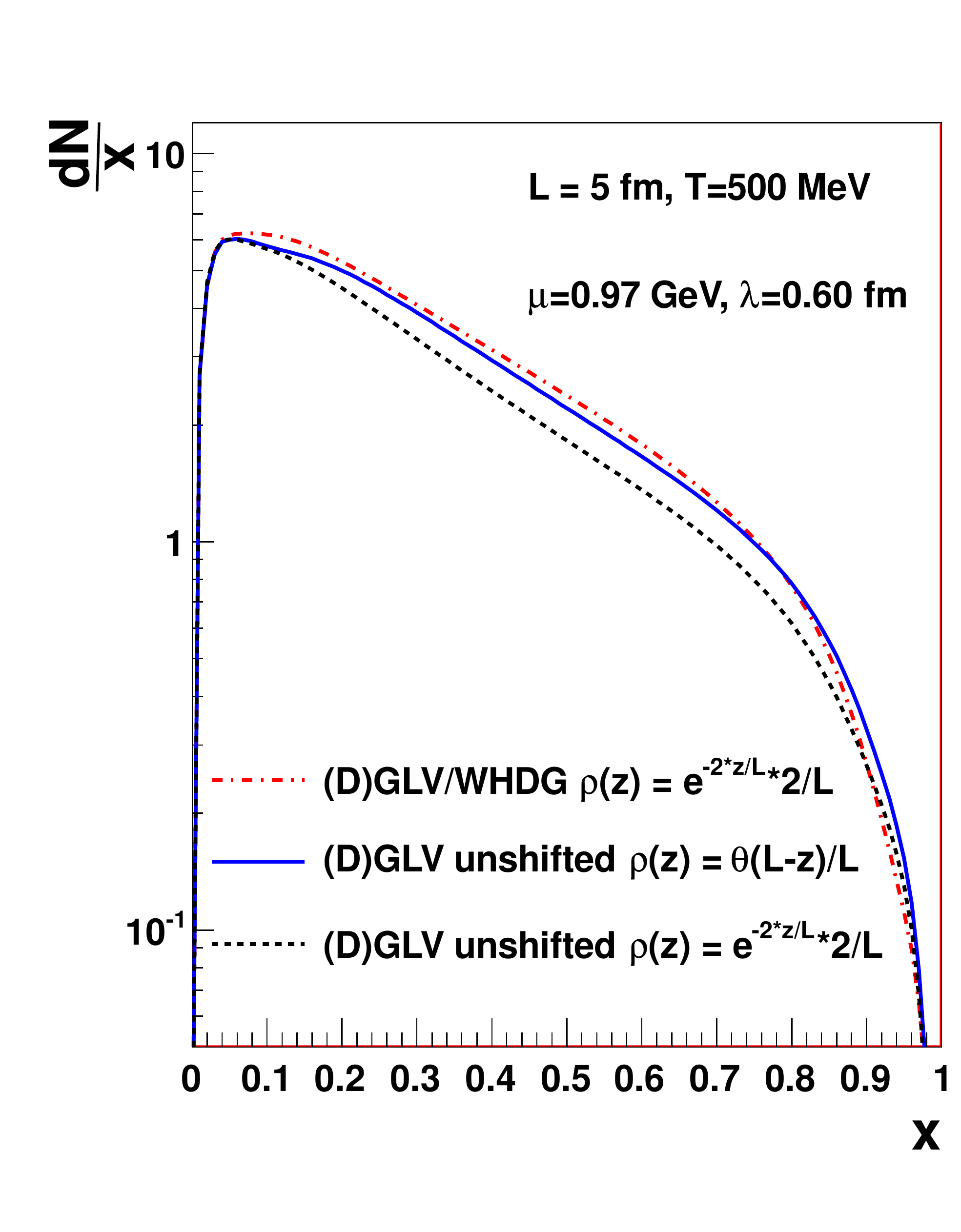}}
    \hspace{.03in} \subfigure[Comparison to single gluon spectrum with
      temperature profile following an expanding glauber
      profile.]{\label{fig:SingleGluonSpecTemp}
      \includegraphics[width=0.35\textwidth]{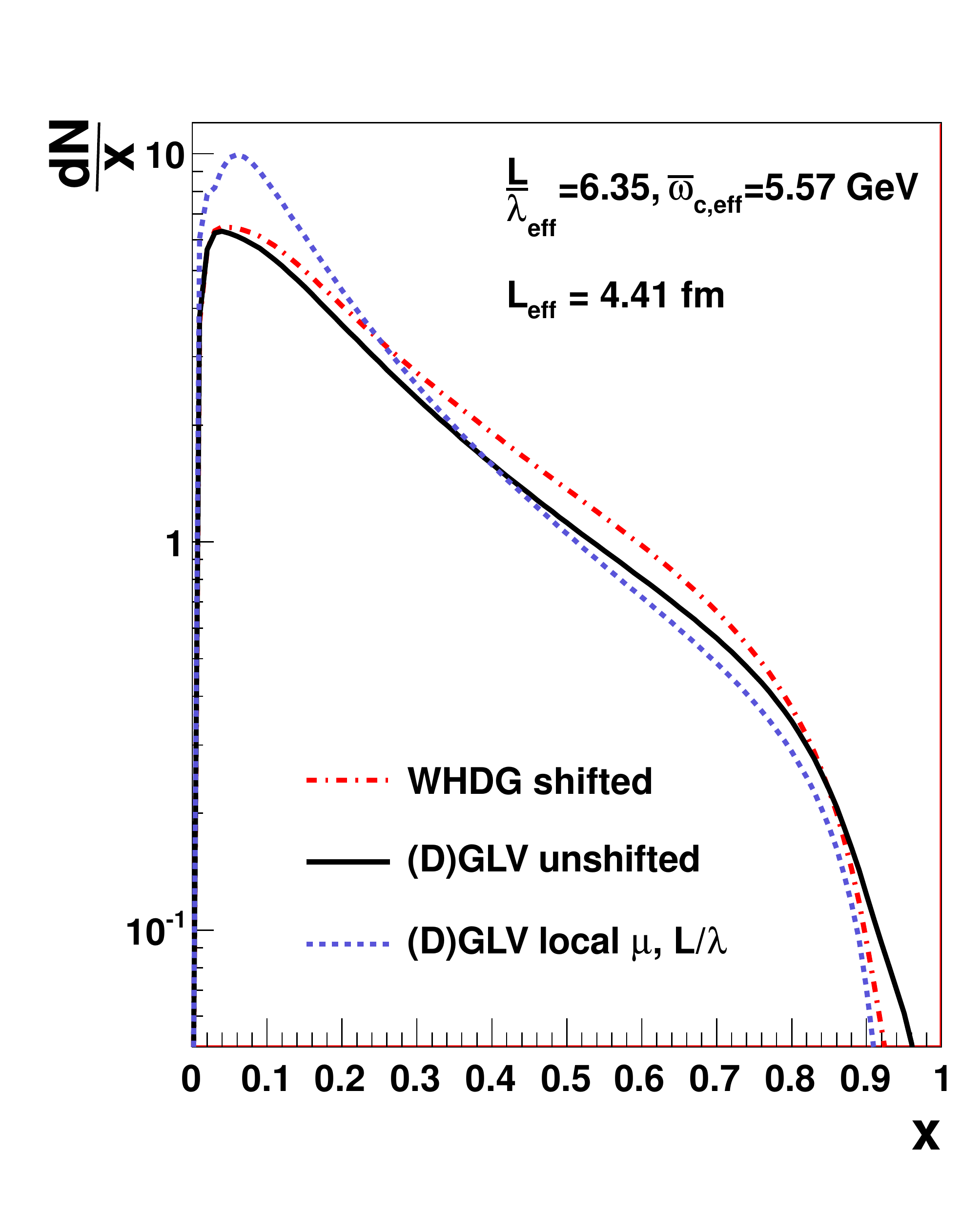}}
    \caption{Single gluon spectra}
    \label{fig:dIdwGLV}
\end{figure*}

In the original GLV implementation the distance of scattering centers is
assumed to be an exponentially decaying distribution, $\rho(z) = \frac{2}{L}
e^{-2z/L}$.  Another widely used choice for the density of scattering centers
is a uniform density (brick): $\rho(z) = \frac{1}{L}\theta(L-z)$ if $z \leq
L$.  Figure \ref{fig:SingleGluonSpec} shows the single gluon emission spectrum
for these two density profiles.  The curve labeled 'shifted' corresponds to
the WHDG calculation \cite{Wicks:2005gt}, which uses a variable shift {\bf q
  $\rightarrow$ q + k} and the approximation {\bf k} $\gg$ {\bf q} for
analytical convenience, for more details see \cite{Horowitz:2009eb}. In figure
\ref{fig:SingleGluonSpec} we see that the three curves show a similar
radiation spectrum. Within the GLV formalism the total energy loss does not
strongly depend on the exact shape of $\rho(z)$.

\subsection{Temperature dependence}
In figure \ref{fig:SingleGluonSpecTemp} we show an additional curve in which
the scattering rate of equation \ref{eq:Cqz} depends on the local temperature
of the medium. We implement a Bjorken-expanding-Glauber profile and let all
temperature dependent variables evolve following this profile. The parton
starts in the center of the medium and moves radially outwards. The result is
compared with the calculation based on an effective medium density, which is
calculated using equation \ref{eq:effGLV}: $L/\lambda = 0.35$,
$\bar{\omega_{c}} = 5.57$ GeV and $L = 4.41$ fm.
The position dependent temperature profile enters the scattering rate via the
density $\mathcal{N}(z) =
\frac{\zeta(3)}{\zeta(2)}(1+\frac{1}{4}N_{f})T^{3}(z)$ \cite{Arnold:2008vd}.

\begin{equation}\label{eq:GammaComb}
\frac{d\Gamma_{el}}{d^{2}q_{\perp}} \simeq \frac{C_{R}}{(2\pi)^{2}}C({\bf q},z)  \simeq \frac{C_{R}}{(2\pi)^{2}}
  \times \frac{g^{4}\mathcal{N}(z)}{(q_{\perp}^{2}+\mu^{2}(z))^{2}},
\end{equation}
in which $C_{R}$ is the Casimir factor; $g=\sqrt{4\pi\alpha_{s}}$;
$\mu = \sqrt{1+\frac{1}{6}N_{f}}\sqrt{4\pi\alpha_{s}}T$, the Debeye screening mass.
The temperature $T$ is the local temperature of the medium and depends
on the position $z$ of the parton in the medium.
Taking into account the local temperature of the medium there is more soft
gluon radiation compared to the corresponding brick.

\begin{figure}[!h]
\centering
\includegraphics[width=0.5\textwidth]{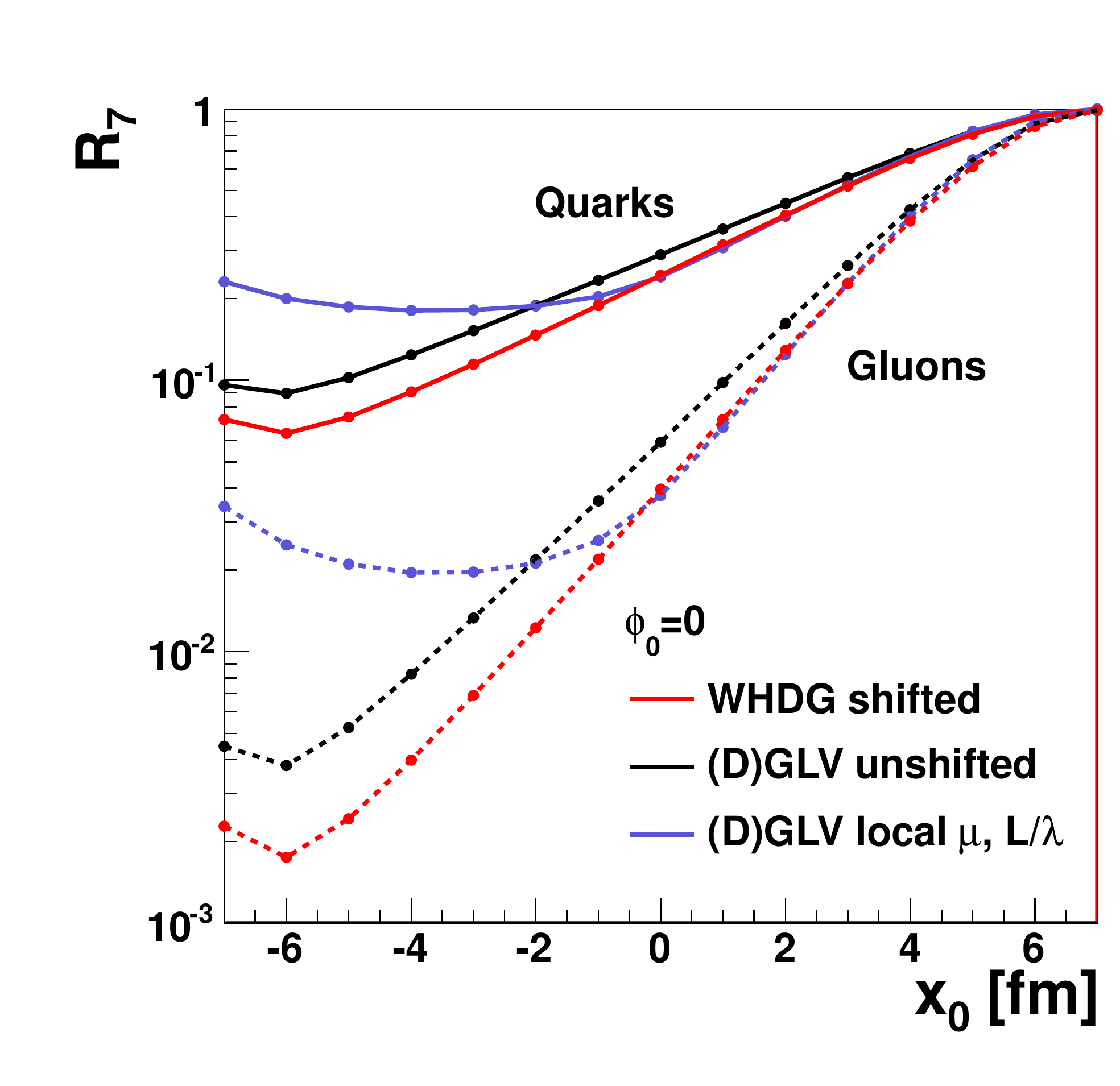}
\caption{Suppression factor, $R_{7}$, as function of the starting point
  $x_{0}$ of parton for temperature dependent distribution of scattering
  centers and equivalent brick \label{fig:R7vsx0}}
\end{figure}

In figure \ref{fig:R7vsx0} we show the path length dependence of suppression
factor $R_{n} = \int_{0}^{1}d\epsilon (1-\epsilon)^{n-1}P(\epsilon)$ which is
an approximation for $R_{AA}$ in a brick. The fraction of the radiated energy
by multiple gluon emissions is $\epsilon$, $P(\epsilon)$ is the
energy loss probability distribution and $n=7$ for RHIC energies. The parton
is created at $x_{0}$ and travels in the positive direction on the x-axis. The
suppression factor is given as function of the starting position of the
parton. For short path lengths the temperature dependent description agrees
with the corresponding brick. For large path lengths however the suppression
of the brick is much larger than for the temperature dependent calculation.

While using a common description of the density for different description of
the energy loss in a hot QCD medium (ASW-MS, ASW-SH and (D)GLV) the estimated
$\hat{q}$ differs up to a factor of $5$. In addition we conclude that using
average medium parameters for an evolving medium overestimates the energy loss
for long parton trajectories.



\bibliographystyle{elsarticle-num}
\bibliography{biblioHP2010}

\begin{thebibliography}{10}
\expandafter\ifx\csname url\endcsname\relax
  \def\url#1{\texttt{#1}}\fi
\expandafter\ifx\csname urlprefix\endcsname\relax\def\urlprefix{URL }\fi
\expandafter\ifx\csname href\endcsname\relax
  \def\href#1#2{#2} \def\path#1{#1}\fi

\bibitem{Salgado:2003gb}
C.~A. Salgado, U.~A. Wiedemann, {Calculating quenching weights}, Phys. Rev. D68
  (2003) 014008.
\newblock \href {http://arxiv.org/abs/hep-ph/0302184}
  {\path{arXiv:hep-ph/0302184}}, \href
  {http://dx.doi.org/10.1103/PhysRevD.68.014008}
  {\path{doi:10.1103/PhysRevD.68.014008}}.

\bibitem{Gyulassy:1999zd}
M.~Gyulassy, P.~Levai, I.~Vitev, {Jet quenching in thin quark-gluon plasmas. I:
  Formalism}, Nucl. Phys. B571 (2000) 197--233.
\newblock \href {http://arxiv.org/abs/hep-ph/9907461}
  {\path{arXiv:hep-ph/9907461}}, \href
  {http://dx.doi.org/10.1016/S0550-3213(99)00713-0}
  {\path{doi:10.1016/S0550-3213(99)00713-0}}.

\bibitem{Wicks:2005gt}
S.~Wicks, W.~Horowitz, M.~Djordjevic, M.~Gyulassy, {Elastic, Inelastic, and
  Path Length Fluctuations in Jet Tomography}, Nucl. Phys. A784 (2007)
  426--442.
\newblock \href {http://arxiv.org/abs/nucl-th/0512076}
  {\path{arXiv:nucl-th/0512076}}, \href
  {http://dx.doi.org/10.1016/j.nuclphysa.2006.12.048}
  {\path{doi:10.1016/j.nuclphysa.2006.12.048}}.

\bibitem{Dainese:2004te}
A.~Dainese, C.~Loizides, G.~Paic, {Leading-particle suppression in high energy
  nucleus nucleus collisions}, Eur. Phys. J. C38 (2005) 461--474.
\newblock \href {http://arxiv.org/abs/hep-ph/0406201}
  {\path{arXiv:hep-ph/0406201}}, \href
  {http://dx.doi.org/10.1140/epjc/s2004-02077-x}
  {\path{doi:10.1140/epjc/s2004-02077-x}}.

\bibitem{Kniehl:2000fe}
B.~A. Kniehl, G.~Kramer, B.~Potter, {Fragmentation functions for pions, kaons,
  and protons at next-to-leading order}, Nucl. Phys. B582 (2000) 514--536.
\newblock \href {http://arxiv.org/abs/hep-ph/0010289}
  {\path{arXiv:hep-ph/0010289}}, \href
  {http://dx.doi.org/10.1016/S0550-3213(00)00303-5}
  {\path{doi:10.1016/S0550-3213(00)00303-5}}.

\bibitem{Adare:2008cg}
A.~Adare, et~al., {Quantitative Constraints on the Opacity of Hot Partonic
  Matter from Semi-Inclusive Single High Transverse Momentum Pion Suppression
  in Au+Au collisions at $\sqrt{s_{NN}}$ = 200 GeV}, Phys. Rev. C77 (2008)
  064907.
\newblock \href {http://arxiv.org/abs/0801.1665} {\path{arXiv:0801.1665}},
  \href {http://dx.doi.org/10.1103/PhysRevC.77.064907}
  {\path{doi:10.1103/PhysRevC.77.064907}}.

\bibitem{Adare:2008qa}
A.~Adare, et~al., {Suppression pattern of neutral pions at high transverse
  momentum in Au+Au collisions at $\sqrt{s_{NN}}$ = 200 GeV and constraints on
  medium transport coefficients}, Phys. Rev. Lett. 101 (2008) 232301.
\newblock \href {http://arxiv.org/abs/0801.4020} {\path{arXiv:0801.4020}},
  \href {http://dx.doi.org/10.1103/PhysRevLett.101.232301}
  {\path{doi:10.1103/PhysRevLett.101.232301}}.

\bibitem{CaronHuot:2010bp}
S.~Caron-Huot, C.~Gale, {Finite-size effects on the radiative energy loss of a
  fast parton in hot and dense strongly interacting matter}\href
  {http://arxiv.org/abs/1006.2379} {\path{arXiv:1006.2379}}.

\bibitem{Horowitz:2009eb}
W.~A. Horowitz, B.~A. Cole, {Systematic theoretical uncertainties in jet
  quenching due to gluon kinematics}, Phys. Rev. C81 (2010) 024909.
\newblock \href {http://arxiv.org/abs/0910.1823} {\path{arXiv:0910.1823}},
  \href {http://dx.doi.org/10.1103/PhysRevC.81.024909}
  {\path{doi:10.1103/PhysRevC.81.024909}}.

\bibitem{Arnold:2008vd}
P.~Arnold, W.~Xiao, {High-energy jet quenching in weakly-coupled quark-gluon
  plasmas}, Phys. Rev. D78 (2008) 125008.
\newblock \href {http://arxiv.org/abs/0810.1026} {\path{arXiv:0810.1026}},
  \href {http://dx.doi.org/10.1103/PhysRevD.78.125008}
  {\path{doi:10.1103/PhysRevD.78.125008}}.

\end{thebibliography}







\end{document}